\documentclass[lettersize,journal]{IEEEtran}
\usepackage{algorithmic}
\usepackage{algorithm}
\usepackage{orcidlink}

\usepackage{multirow}
\usepackage{cite}
\usepackage{svg}
\usepackage[nolist,nohyperlinks]{acronym}
\begin{acronym}[ICANN]

\begin{acronym}[]
    \acro{IoT}{internet of things}
    \acro{IIoT}{industrial internet of things}
    \acro{UAV}{unmanned aerial vehicle}
    \acro{AGV}{autonomous guided vehicle}
    \acro{IPS}{indoor positioning system}
    \acro{SOTA} {state-of-the-art}

    \acro{RF}{radio frequency}
    \acro{PRF}{pulse repetition frequency}
    \acro{VLP}{visible light positioning}
    \acro{MOCAP}{motion capturing}
    \acro{UWB}{ultra-wideband}
    \acro{TWR}{two-way-ranging}
    \acro{SS-TWR}{single sided two-way-ranging}
    \acro{DS-TWR}{double sided two-way-ranging}
    \acro{ADS-TWR}{asymmetric double sided two-way-ranging}
    \acro{SDS-TWR}{symmetric double sided two-way-ranging}
    \acro{TDoA}{time-difference-of-arrival}
    \acro{DDoA}{distance-difference-of-arrival}
    \acro{AoA}{angle-of-arrival}
    \acro{RSSi}{receive signal strength indicator}
    \acro{ToF}{time-of-flight}
    \acro{PF}{particle filter}
    \acro{CIR}{Channel Impulse Response}
    \acro{CFO}{carrier frequency offset}
    \acro{CFAR}{constant false alarm rate}
    \acro{CA-CFAR}{cell averaging \ac{CFAR}}
    \acro{PBR}{phase-based ranging}
    
    \acrodef{ADC}[ADC]{analog-to-digital converter}
    \acro{SLAM}{Simultaneous Localization and Mapping}
    \acro{AGV}{autonomous guided vehicle}
    \acro{NLOS}{non-line-of-sight}
    \acro{LOS}{line-of-sight}
    
    \acro{ML}{machine learning}
    \acro{CNN}{convolutional neural network}
    \acro{NN}{neural network}
    \acro{TL}{transfer learning}

    \acro{PEKF}{periodic extended Kalman filter}
    \acro{EKF}{extended Kalman filter}
    \acro{LS}{least-squares}
    \acro{PF}{particle filter}
    \acro{AF}{averaging filter}
    \acro{SG}{Savitzky-Golay}
    \acro{EC}{error correction}

    \acro{MAE}{mean absolute error}
    \acro{AE}{absolute error}
    \acro{SE}{spatial error}
    \acro{MSE}{mean spatial error}
    \acro{MPC}{multipath components}
    \acro{cdf}{cumulative distribution function}

    \acro{IPIN}{Indoor Positioning and Indoor Navigation}
    \acro{IPSN}{Information Processing in Sensor Networks}

    \acro{BMA}{bi-static mean algorithm}
    \acro{BSDA}{bi-static standard deviation algorithm}
    \acro{MMA}{multi-static mean algorithm}
    \acro{MSDA}{multi-static superposition algorithm}
    \acro{COTS}{commercial off-the-shelf}

    \acro{ICIR}{integer channel impulse response}
    \acro{UCIR}{upsampling channel impulse response}
    \acro{ACIR}{accumulation channel impulse response}
\end{acronym}
\end{acronym}

\usepackage{empheq}
\usepackage{amsmath,amsfonts}
\usepackage{algorithmic}
\usepackage{array}
\usepackage[caption=false,font=normalsize,labelfont=sf,textfont=sf]{subfig}
\usepackage{textcomp}
\usepackage{stfloats}
\usepackage{url}
\usepackage{verbatim}
\usepackage{graphicx}
\def\BibTeX{{\rm B\kern-.05em{\sc i\kern-.025em b}\kern-.08em
    T\kern-.1667em\lower.7ex\hbox{E}\kern-.125emX}}
\usepackage{balance}

\begin{document}

\title{Towards mm-Level Accurate UWB Radar: High-Accuracy Phase-Based Obstacle Detection through Multi-Channel Fusion}

\author{Jelle De Moerloose \orcidlink{0009-0006-8061-3692}, Adnan Shahid \orcidlink{0000-0003-1943-6261}, Eli De Poorter \orcidlink{0000-0002-0214-5751}
\thanks{The authors are with IDLab, Department of Information Technology, Ghent University-imec, Ghent, Belgium (e-mail: Jelle.DeMoerloose@UGent.be).}
}

\markboth{Submitted to IEEE Transactions on Wireless Communications, June~2026}%
{De Moerloose \MakeLowercase{\textit{et al.}}: Towards mm-Level Accurate UWB Radar: High-Accuracy Phase-Based Obstacle Detection through Multi-Channel Fusion}

\IEEEpubid{0000--0000/00\$00.00~\copyright~2026 IEEE}

\maketitle
\begin{abstract}
Accurate, tag-free distance estimation with ultra-wideband (UWB) radar is essential for applications such as autonomous guided vehicles, robotics, and environment characterization. For tag-based localization systems, phase-based UWB signal processing techniques have demonstrated sub-wavelength ranging precision, but these approaches are not applicable for passive (tagless) radar setups with weak reflections, mixed multipath conditions, and the absence of a known \acf{ToF} first-path reference. This paper demonstrates for the first time that phase information can be effectively exploited in a fully passive UWB radar setting. We introduce a signal processing framework that extracts reliable distance information by combining coarse amplitude-based estimates with high-resolution phase changes across multiple frequency channels. By referencing phase measurements with the line-of-sight component, the method compensates for hardware-induced phase drift, while the use of multi-channel frequency diversity enables disambiguation of periodic phase information and improves robustness against frequency-specific channel degradation such as Fresnel zones. The proposed approach is validated on a robot equipped with a bistatic UWB radar using DW3000 devices and evaluated in a realistic metallic industrial environment. Experimental results show that our work consistently achieves centimeter-level accuracy even at high speeds, with a median error of 1.69 cm, significantly outperforming existing $\sim$10cm accuracy UWB radar approaches relying only on amplitude-information. We further show how multi-channel fusion exploits uncorrelated channel degradation to reduce the error by more than 40\% compared to single-channel operation, and outline how phase modeling and fusion can be pushed toward sub-centimeter accuracy.

\end{abstract}


\section{Introduction}
\IEEEPARstart{A}{ccurate} and scalable distance estimation is a fundamental requirement for emerging applications requiring precise positioning and environmental perception. Applications such as collision avoidance for \ac{AGV}\cite{AVG}, SLAM\cite{SLAM}, VR/AR spatial tracking \cite{VR}, environment mapping \cite{environment_mapping}, handwriting detection \cite{ma2024push, hand_writing}, and vital sign monitoring \cite{lambrecht2025lowcostembeddedbreathingrate} all depend on highly precise distance measurements to operate reliably. For many of these applications, tagless (passive) operation is desirable, as attaching dedicated tags to objects of interest reduces scalability and deployment flexibility.

\Ac{UWB} technology is a promising candidate for such applications due to its large bandwidth (0.5 to 2.5 GHz) which enables fine temporal resolution and the separation of multipath components in the \ac{CIR}. Unlike alternatives such as LiDAR, camera, or mmWave, it offers fine-grained ranging without sacrificing privacy, energy efficiency, or cost.

Despite these advantages, \acf{ToF} UWB ranging remains limited to centimeter-level accuracy (5–10 cm) in ideal conditions \cite{TWR_ranging_Acc, XRLoc}, insufficient for fine-grained motion capture. Recently, Ma et al. \cite{ma2024push} and UTrack3D \cite{UTrack3D} broke this barrier by exploiting the phase of the received signal, which encodes sub-wavelength distance information beyond what amplitude-based \ac{ToF} can resolve, achieving mm-level ranging and tracking accuracy. However, both rely on a tag attached to the target. To the best of our knowledge, phase information has never been applied for tagless UWB radar, where both amplitude-based distance estimation and phase extraction become fundamentally harder.

For amplitude-based distance estimation, reflected signals in the CIR are significantly weaker and arrive mixed with multipath, requiring robust peak selection from multiple low-SNR reflections. Commercial 802.15.4 UWB hardware compounds this by providing high-resolution timing only for the first path, leaving reflected paths at a much coarser native CIR resolution. Using phase information adds three further challenges. (i) Phase measurements drift over time due to hardware-induced offsets and unsynchronized oscillators, making them unstable without active correction. (ii) Phase observations are inherently ambiguous, as multiple propagation distances map to the same wrapped phase. (iii) The absence of a dominant reference reflection produces noisier and less informative phase measurements than tag-based systems, since the measured phase is a superposition of multiple weak components. Together, these factors have so far prevented phase-based techniques from being applied in tagless UWB radar.

\IEEEpubidadjcol
This paper addresses all of the above challenges with the following main contributions:

\begin{itemize}
\item Passive phase-based UWB sensing: We show that phase information, previously limited to tag-based systems, can be effectively utilized for high-accuracy distance estimation in passive UWB radar on commercial hardware.
\item Low-complexity phase drift cancellation: We design a LOS-referenced method that cancels phase offset and drift in unsynchronized bistatic configurations, enabling stable phase extraction over time.
\item Multi-channel \ac{PF} fusion: We fuse amplitude and phase across multiple UWB channels inside a \ac{PF}, using spectral diversity to resolve phase ambiguity and to remain robust to channel-specific degradation.
\item Experimental validation in realistic conditions: We validate the approach on a mobile UWB radar platform in a metallic industrial environment, achieving 1.69 cm median error and robustness across ambiguous motion speeds.
\end{itemize}

\noindent The remainder of this paper is organized as follows. Section~\ref{sec:related} reviews related work on high-accuracy UWB ranging. Section~\ref{sec:system_model} presents the system model and formulates the passive phase-based ranging problem. Section~\ref{sec:methodology} details the proposed multi-channel fusion methodology. Section~\ref{sec:implementation} describes the system implementation. Section~\ref{sec:evaluation} evaluates the proposed approach through experiments in a realistic industrial environment. Finally, Section~\ref{sec:future_work} proposes directions for future work and Section~\ref{sec:conclusion} concludes the paper.
\section{Related work}
\label{sec:related}
\noindent Several prior works have explored high-precision UWB distance estimation, although none achieve sub-centimeter accuracy in a fully passive setting. We first review amplitude-based passive radar, the dominant paradigm on hardware, and then phase-based ranging, which achieves the highest reported accuracies but relies on cooperative tags. Table~\ref{tab:related_work} summarizes the most representative studies, detailing their techniques, key characteristics, and achieved accuracy.
 
\begin{table*}[t]
\caption{Comparison of high-accuracy UWB ranging and radar approaches, summarizing their techniques, robustness characteristics, and achieved accuracy for both tag-based and passive systems.}
\label{tab:related_work}
\centering
\footnotesize
\begin{tabular}{| l | l | p{3cm} | c | c | p{3cm} | c |}
\hline
\textbf{Category} & \textbf{Paper} & \textbf{Technique} & \textbf{Phase} & \textbf{Robust to Ch.} & \textbf{Robot speed} & \textbf{Accuracy} \\
 & & & \textbf{Used} & \textbf{Degradation} & \textbf{(Theoretical limit)} & \\
\noalign{\hrule height 1.5pt}
\begin{tabular}[t]{@{}l@{}}\textbf{Tag-based} \\ \textbf{ranging}\end{tabular}
& Ma et al.~\cite{ma2024push}
& Phase-based dual-frequency ranging
& $\checkmark$ & $\times$
& \begin{tabular}[t]{@{}l@{}} Tested to 0.5 m/s \\ \scriptsize{(Duration of TWR operations)} \end{tabular}
& $<$1 mm MedAE \\
\cline{2-7}
& UTrack3D~\cite{UTrack3D}
& Phase-difference relative tracking
& $\checkmark$ & $\times$
& \begin{tabular}[t]{@{}l@{}} Degrades $>$0.65 m/s \\ \scriptsize{(Phase ambiguity)} \end{tabular}
& 9 mm (P90) \\
\noalign{\hrule height 1.5pt}
\begin{tabular}[t]{@{}l@{}}\textbf{Tagless} \\ \textbf{radar}\end{tabular}
& Van Herbruggen et al.~\cite{CoarseDistance}
& ACIR with BMA
& $\times$ & $\times$
& \begin{tabular}[t]{@{}l@{}} Static only \\ \scriptsize{(CIR accumulation)} \end{tabular}
& $<$9 cm MAE \\
\cline{2-7}
& Giurea et al.$^\dagger$~\cite{Adelina}
& Peak detection with filtering criteria
& $\times$ & $\times$
& \begin{tabular}[t]{@{}l@{}} Tested to 1.7 m/s$^\dagger$ \\ \scriptsize{(N/A)} \end{tabular}
& 8.44 cm MedAE \\
\cline{2-7}
& \textbf{This work}
& \textbf{Multi-channel ToF \& Phase fusion (PF)}
& $\checkmark$ & $\checkmark$
& \begin{tabular}[t]{@{}l@{}} \textbf{Tested to 1.7 m/s} \\ \scriptsize{\textbf{(No degradation observed)}} \end{tabular}
& \textbf{1.69 cm MedAE} \\
\hline
\end{tabular}
\vspace{1mm}
{\footnotesize $^\dagger$Recreated as a baseline implementation and evaluated on our dataset.}
\end{table*}
 
\subsection{Amplitude-Based UWB Radar}
\label{section:cir_processing}
\noindent Amplitude-based pipelines combine three processing stages to surpass the coarse native resolution of the \ac{CIR}, roughly 30~cm per bin for a 500~MHz bandwidth (Section~\ref{accumulation_explanation}). Preprocessing improves CIR quality through smoothing \cite{Li2025}, background removal \cite{base_signal_proccess,Yun2019}, and interpolation \cite{intrapolation} or accumulation \cite{ACIR}. Peak detection then identifies the true object reflection, where \ac{CFAR}-based methods adaptively set thresholds against local noise \cite{CFAR_1, CFAR_2, CFAR_3} and CLEAN-based approaches iteratively remove multipath artifacts \cite{base_signal_proccess}. Post-processing exploits temporal correlations to refine the estimates, using Kalman filtering \cite{SLAM,cfkalman}, particle filtering \cite{particle_filtering,DL2}, or deep learning \cite{DL_1,DL2}. Even with such advanced processing, reported passive amplitude-based accuracies remain at decimeter level, insufficient for fine-grained applications.
 
The work closest to this accuracy ceiling is Van Herbruggen et al.~\cite{CoarseDistance}, who explored different CIR processing methods in a bistatic passive radar setup. They reported state-of-the-art accuracy ($<$9~cm MAE) using a CIR accumulation strategy (ACIR), which averages multiple CIR samples aligned by the reported first-path index. However, this method requires both the radar and the target to remain stationary during the accumulation phase, limiting its applicability in real-world scenarios. Moreover, their peak detection relies on background subtraction, making it unsuitable for a moving robot where the environment is constantly changing.
 
Most directly comparable to this work, Giurea et al.~\cite{Adelina} propose a passive UWB obstacle detection pipeline on a DW3000-equipped mobile robot, using the same hardware and deployment context as this work. Their peak detection is based on width, prominence, and SNR thresholding, similar in spirit to standard \ac{CA-CFAR}~\cite{UWB_overview}. Unlike our approach, it relies solely on amplitude information without phase-based refinement. Their pipeline additionally performs AoA filtering and 2D clustering, which requires multi-antenna recordings unavailable in our setup. We implement their peak detection algorithm and evaluate it as a baseline in Section~\ref{section:baseline}.
 
\subsection{Phase-Based Tag Ranging}
\begin{figure}[t]
    \centering
    \begin{minipage}[c]{0.5\textwidth}
        \centering
        \includegraphics[width=\textwidth]{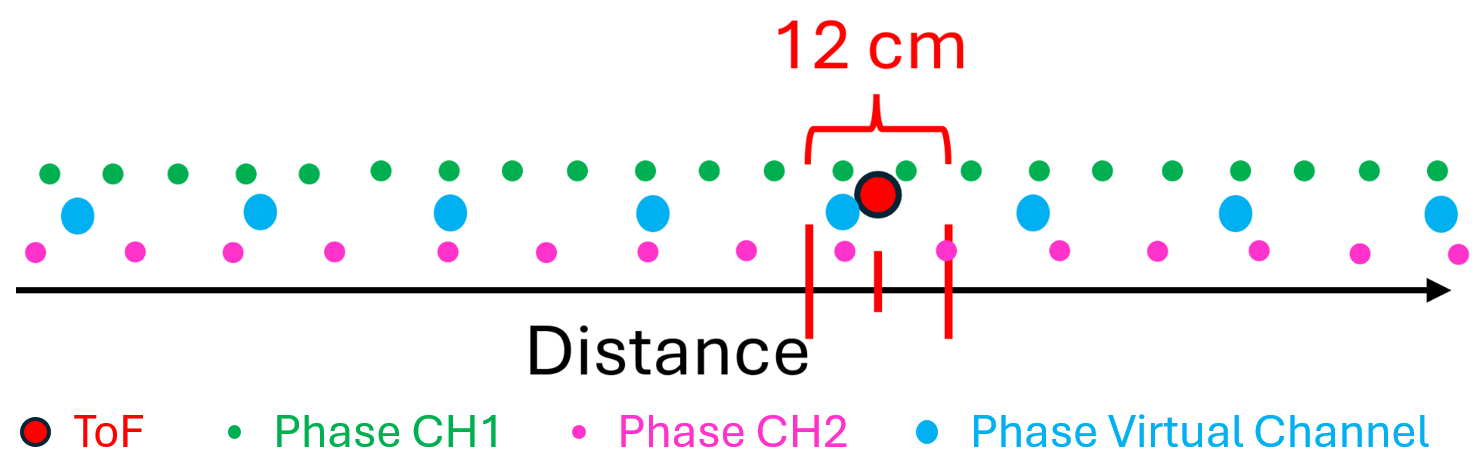}
    \end{minipage}%
 
    \caption{Illustration of the dual-channel phase combination method proposed by Ma et al. \cite{ma2024push}, where phase measurements are combined into a virtual channel to resolve ambiguity within a narrow amplitude-based ToF uncertainty window. This approach does not directly translate to passive radar, where the uncertainty window is significantly larger and individual channels may be unreliable, making virtual channel construction insufficient. }
    \label{fig:virtual_channel}
\end{figure}

\noindent Ma et al.~\cite{ma2024push} pushed the limits of tag-based ranging by being the first to exploit CIR phase information, achieving exceptional sub-millimeter accuracies. The carrier phase repeats every wavelength, so the phase of a single channel cannot resolve an absolute position within the 12~cm uncertainty window of their amplitude-based \ac{ToF} estimate, as illustrated in Fig.~\ref{fig:virtual_channel}. They resolve this ambiguity by subtracting the phases of two channels, creating a virtual channel with an inherently larger wavelength, and refine the final estimate using only the highest frequency channel, since the virtual channel carries more noise. To obtain usable phase in the first place, they cancel phase drift through an extended \ac{TWR} protocol with channel switching.
 
Three limitations make this method unsuitable for UWB radar. (i) It relies on the high-accuracy first path, which provides a small ToF uncertainty window and high-quality phase measurements. Radar conditions with low-SNR and multipath reflections typically violate both. (ii) It requires reliable measurements on both channels, since channel failure or interference on a single channel directly corrupts the virtual channel and leads to incorrect range estimates. (iii) Each distance estimate requires two complete TWR operations, which fundamentally limits the achievable update rate and implicitly assumes a quasi-constant transmitter-receiver distance throughout the message exchange. This directly limits supported target speeds, which their evaluation confirms by testing only up to 0.5~m/s.
 
UTrack3D \cite{UTrack3D} exploits single-channel phase information to achieve a reported 9~mm tracking accuracy at the 90th percentile. After multi-antenna phase cancellation, phase differences between consecutive measurements estimate relative displacement. Using three spatially separated antennas, the system tracks 3D motion starting from an initial position obtained via TWR, after which tracking relies purely on relative phase-based updates. The inferred displacement is ambiguous when inter-sample motion exceeds half the carrier wavelength, giving a theoretical maximum speed of approximately 1.45~m/s. In practice, phase noise causes wrapping errors to emerge well before this limit. Moreover, the reported accuracy reflects relative tracking error, as trajectories are aligned to ground truth at initialization rather than evaluating absolute localization error. Errors due to phase noise or channel degradation accumulate over time, making the approach sensitive to channel failures and requiring periodic re-initialization.
 
To the best of our knowledge, Ma et al. \cite{ma2024push} and UTrack3D \cite{UTrack3D} are among the first works to exploit CIR phase information from commercial UWB devices for high-accuracy ranging and tracking, though both depend on a small ToF uncertainty window, reliable channels, and high-quality phase information. This work is the first to exploit the CIR phase without these assumptions, in a tagless radar setting and leveraging multi-channel diversity for improved accuracy and robustness against channel degradation.
 
\begin{figure*}[t]
    \centering
    \includegraphics[width=0.85\textwidth]{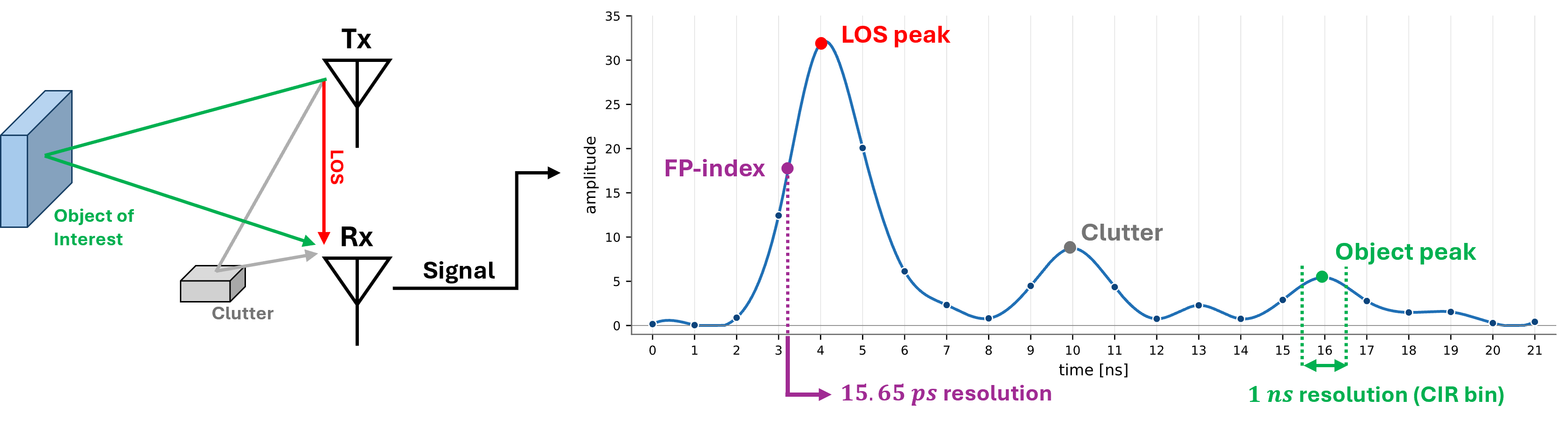}
    \caption{ Bistatic radar scene (left) and measured CIR (right). The strong LOS peak and its hardware-provided FP-index at 15.65 ps resolution carry highly accurate range information in tag-based operation, where such a peak originates from the target. In the passive case, this describes the Tx-Rx path instead, and the object reflection must be recovered through peak detection across 1 ns CIR bins (30 cm resolution), where it appears weak and mixed with clutter.}
    \label{fig:radarvstag}
\end{figure*}
 
\section{System Model and Problem Formulation}
\label{sec:system_model}
\noindent This section models the observations available to a passive UWB radar on commercial hardware and formalizes the estimation problem this paper addresses.
 
\subsection{Signal Model}
\label{accumulation_explanation}
\noindent A UWB transmitter emits a signal consisting of many pulses of only a few nanoseconds, which reflect off objects and arrive at the receiver via multiple paths. The receiver captures this propagation behavior in the \acf{CIR}, which contains all paths with their respective delays, amplitudes, and phases:
\begin{equation}
h(t) = \sum_{k=0}^{K-1} \alpha_k e^{j\phi_k} \delta(t - \tau_k) + n(t)
\label{eq:CIR}
\end{equation}
Here, $K$ is the number of multipath components, $\alpha_k$ and $\phi_k$ are the amplitude and phase of the $k$-th path, $\tau_k$ is its delay, and $n(t)$ is complex Gaussian noise. As such, the CIR provides a time-resolved view of the environment and forms the basis for \ac{ToF} ranging.
 
In hardware such as the DW3000, the short pulse duration (e.g., 2~ns for 500~MHz) makes direct sampling challenging \cite{qorvo_dw3000_datasheet}. The CIR is instead reconstructed via a correlation-based approach: a known preamble is transmitted, and the receiver correlates the incoming signal with it. The correlation outputs are accumulated across multiple symbols, forming discrete CIR bins that represent the combined contribution of all multipath components arriving within each time interval (see Fig.~\ref{fig:radarvstag}) \cite{qorvo_api}. This enables high temporal resolution without ultra-fast \acp{ADC} and improves the signal-to-noise ratio.
 
Each capture thus produces one discrete CIR, a sequence of complex bins like the one shown in Fig.~\ref{fig:radarvstag}. We denote it $h_i^{(c)}[x]$, where the fast-time index $x$ selects a range bin within the CIR and maps to propagation delay, the slow-time index $i$ counts successive captures and maps to measurement time, and $c \in \mathcal{C}$ identifies the UWB channel on which the capture was made. The system observes the environment through these three axes: position along the CIR, evolution over time, and frequency diversity across channels.
 
\subsection{Tag-Based Ranging Versus Passive Radar}
\label{section:tag_vs_radar}
\noindent The two operating modes differ in which path carries the target information, as illustrated in Fig.~\ref{fig:radarvstag}. In tag-based ranging, the target carries a transceiver, so the first-arriving path originates from the target itself. Commercial transceivers report the index of this path (FP-index) with a temporal resolution of 15.65 picoseconds through leading-edge detection \cite{qorvo_api}, enabling highly accurate \ac{ToF} estimation. This high-resolution timing is only available for the first-arriving path.
 
In a passive radar, where one or more devices capture reflections from objects in monostatic, bistatic, or multistatic configurations \cite{CoarseDistance}, the first-arriving path is instead the \ac{LOS} signal between transmitter and receiver, which carries no target information. The right side of Fig.~\ref{fig:radarvstag} shows the consequence: the object reflection appears later in the CIR, weak and mixed with clutter, and must be recovered from the discretized correlation samples reported at a native resolution of approximately 1~ns. For a 500~MHz signal, this corresponds to a Rayleigh resolution of approximately 30~cm per CIR bin\footnote{Following the standard radar principle, the Rayleigh resolution can be approximated as $c/(2B)$, where $c$ is the speed of light and $B$ is the signal bandwidth. This represents the theoretical limit assuming the correct CIR bin is selected.}. Both the coarser timing and the weaker reflections make passive distance estimation considerably harder than tag-based ranging.
 
\subsection{Phase Observation Model}
\label{section:pbr}
\noindent The phase of the received signal encodes distance information beyond the amplitude resolution limit. Since the phase varies continuously within a single carrier wavelength, it captures sub-wavelength changes in the signal's propagation path length. For typical UWB carrier frequencies (6-8\,GHz), the corresponding wavelengths are only a few centimeters (3.8-4.6\,cm), so a $10^{\circ}$ phase shift corresponds to less than 1~mm of target displacement in a radar setup, enabling distance estimates with sub-millimeter resolution.
 
The measured phase $\Phi$ of a received signal corresponds to a distance $d$ according to:
\begin{equation}
    d = \left(\frac{\Phi}{2\pi} + N \right) \cdot \lambda
    \label{eq:phase_distance}
\end{equation}
where $\lambda$ is the signal wavelength, and $N$ is an unknown integer representing the number of full wavelengths between transmitter and receiver.
 
This relation is inherently ambiguous because multiple distance candidates satisfy Eq~\ref{eq:phase_distance}. The receiver only observes the phase wrapped to $(-\pi, \pi]$, so a displacement that produces a phase change exceeding $\pi$ becomes indistinguishable from a smaller displacement in the opposite direction. This occurs when the path length change exceeds $\lambda/2$, corresponding to a displacement of $\lambda/4$ in a passive bistatic radar, where a target motion of $\Delta d$ changes the total path length by $2\Delta d$. The maximum unambiguous speed is therefore:
\begin{equation}
    v_{\max} = \frac{\lambda}{4 \cdot \Delta t}
    \label{eq:vmax}
\end{equation}
 
Furthermore, unsynchronized transmitters and receivers introduce a constant offset and time-varying drift, so the path phases $\phi_k$ of Eq~\ref{eq:CIR} contain more than the propagation geometry. The measured contribution of the $k$-th path evolves over time as:
\begin{equation}
    h_k(t) = \alpha_k e^{-j[2\pi f_c \tau_k + 2\pi \Delta f_c t - (\Phi^{\text{tx}} - \Phi^{\text{rx}})]}
\end{equation} 
Yielding a measured phase of:
\begin{equation}
    \Phi(t) = [-2\pi f_c \tau - 2\pi \Delta f_c t + (\Phi^{\text{tx}} - \Phi^{\text{rx}})] \mod 2\pi
    \label{eq:raw_phase}
\end{equation}
Where $f_c$ is the center frequency, $\tau$ is the time-of-flight, $\Delta f_c$ is the \ac{CFO}, and $(\Phi^{\text{tx}} - \Phi^{\text{rx}})$ is a constant phase offset caused by unsynchronized oscillators. \ac{CFO} arises from small differences in the frequencies generated by individual oscillators, due to manufacturing tolerances and temperature variations \cite{Brink2007}. Both effects cause phase drift over time, which can degrade ranging accuracy.
 
\begin{figure}[t]
    \centering
    \begin{minipage}[c]{0.3\textwidth}
        \centering
        \includegraphics[width=\textwidth]{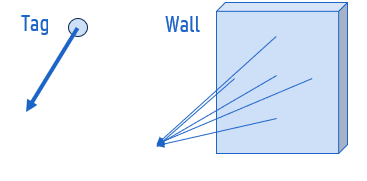}
    \end{minipage}%
 
    \caption{Illustration of a dominant signal from an active UWB tag (thick arrow) compared to weaker multipath reflections from environmental surfaces in a passive radar scenario.}
    \label{fig:dominant_reflection}
\end{figure}
 
Finally, whereas tag-based ranging systems are dominated by a single strong direct path originating from the tag, the phase observed in a passive radar is a composite quantity. Due to the reflection characteristics of larger obstacles, even when the correct peak is selected, the measured phase at that bin is a weighted sum of multiple weak reflections, each with a slightly different path length (Fig.~\ref{fig:dominant_reflection}). The resulting phase behaves like an average and cannot be used directly for high-accuracy absolute ranging. Changes in this superposed phase over time, however, still track the relative motion of the target, a property our methodology exploits.
 
\subsection{Channel Degradation Model}
\label{section:channel_failure}
\begin{figure}[t]
    \centering
    \begin{minipage}[c]{0.5\textwidth}
        \centering
        \includegraphics[width=\textwidth]{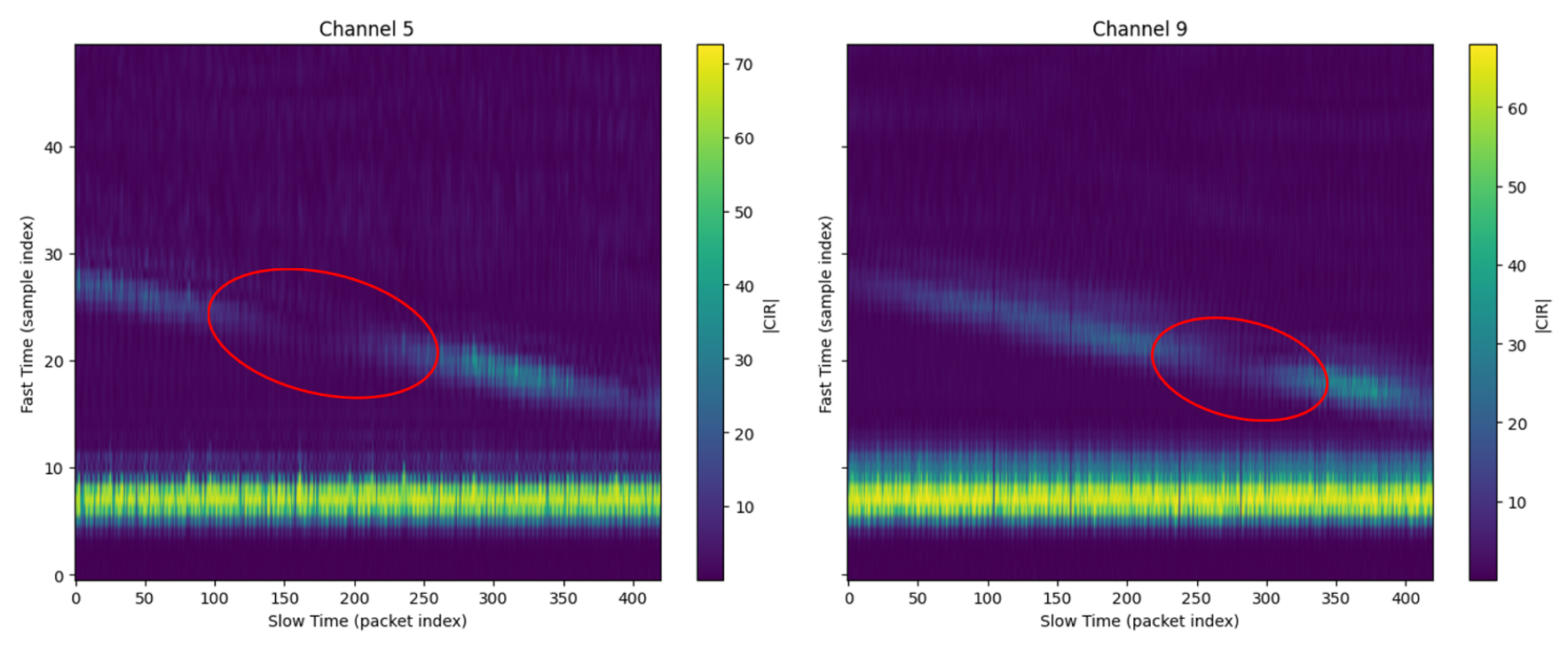}
    \end{minipage}%
 
    \caption{Two CIR heatmaps produced by different channels at the same time having visible interference at different timestamps caused by Fresnel zones. }
    \label{fig:interference}
\end{figure}

\noindent UWB performance can be degraded on a per-channel basis by external interference from co-located wireless systems, or by internal propagation effects such as Fresnel zone obstruction. Fresnel zones describe regions where reflected signal components, most notably from the floor in ground-level robotic deployments, arrive with a path-length difference that causes destructive interference at the direct-path delay, attenuating or distorting the object peak of the \ac{CIR}. This effect is frequency-dependent, as the cancellation geometry changes with the carrier wavelength. Fig.~\ref{fig:interference} illustrates this: two channels recorded simultaneously exhibit interference at different timestamps, making single-channel radar fragile in mobile robotic settings where the geometry changes continuously.
 
\subsection{Problem Formulation}
\label{section:problem_formulation}
\noindent Given the interleaved multi-channel CIR sequences captured from a moving bistatic radar, the goal is to design an estimator $F$ that maps the causal multi-channel measurement history to the scalar distance between the radar and the closest reflection point on the target:
\begin{equation}
\hat{d}_i = F\left(\left\{ h_j^{(c)}[x] \right\}_{j \leq i,\; c \in \mathcal{C}}\right),
\label{eq:estimator}
\end{equation}
such that the absolute ranging error against the ground truth distance $d_i^{\text{GT}}$ is minimized over a trajectory of $T$ measurements:
\begin{equation}
\min_{F} \; \frac{1}{T} \sum_{i=1}^{T} \left| \hat{d}_i - d_i^{\text{GT}} \right|.
\label{eq:objective}
\end{equation}
 
Designing such an estimator faces four obstacles. The target reflection competes with multipath peaks at low SNR, and selecting the wrong CIR bin corrupts both the ToF and the phase extracted at that bin. The wrapped phase of Eq~\ref{eq:phase_distance} maps multiple distances to the same observation, limiting unambiguous single-channel tracking to the speed of Eq~\ref{eq:vmax}. The oscillator terms of Eq~\ref{eq:raw_phase} dominate the raw phase and render it unusable without correction. Finally, the phase at the target bin averages multiple weak reflections, so it cannot serve as an absolute distance reference, and frequency-selective degradation can corrupt individual channels at unpredictable times. Section~\ref{sec:methodology} presents a multi-channel fusion system that addresses these obstacles jointly.

\section{Methodology}
\label{sec:methodology}

\begin{figure*}[t]
    \centering
    \includegraphics[width=0.7\textwidth]{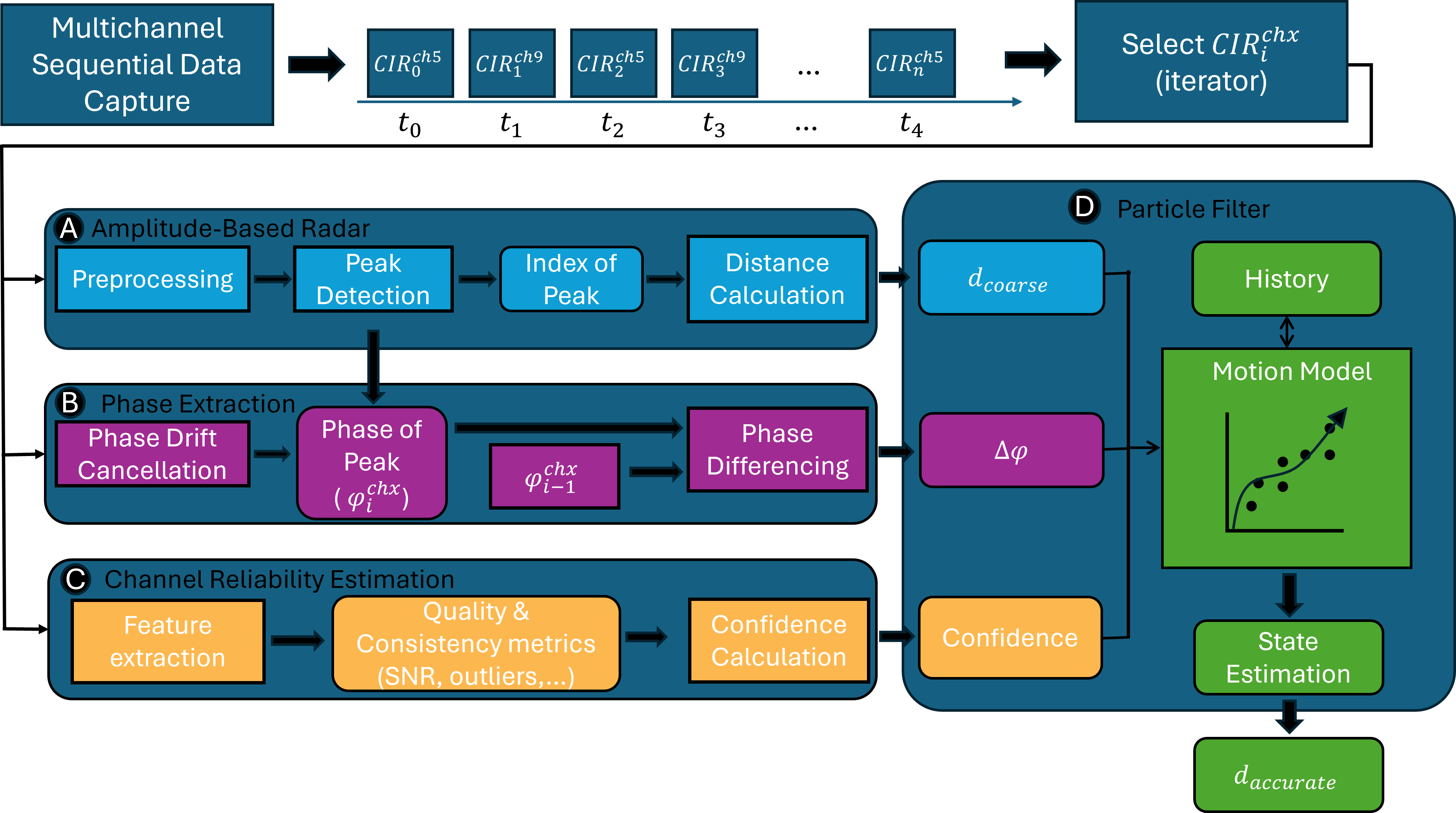}
    \caption{Multichannel processing pipeline for accurate distance estimation. CIR measurements undergo (A) coarse amplitude-based distance estimation, (B) accurate extraction of relative phase changes, and (C) signal reliability estimation, which feed into (D) the \acf{PF} combining motion models and measurement history to produce accurate distance estimates.}
    \label{fig:system overview}
\end{figure*}
\noindent Fig.~\ref{fig:system overview} illustrates the proposed phase-based multi-channel UWB radar system. Raw CIRs are captured sequentially across UWB channels in an interleaved manner, so each channel observes the environment at nearly the same time. Each CIR is fed to three complementary pipelines. The amplitude pipeline (A) extracts the target reflection with \ac{CFAR} and converts its ToF to a coarse absolute distance through the bistatic geometry. The phase pipeline (B) cancels oscillator drift against the always-present \ac{LOS} peak and differences consecutive phase samples to recover sub-bin relative motion, ambiguous by multiples of the carrier wavelength but highly sensitive to displacement. The reliability pipeline (C) maps each measurement's local SNR to a per-channel confidence score, used to suppress phase updates extracted from unreliable peaks.

A particle filter (D) fuses these three streams together with a constant-velocity motion model, jointly resolving distance and velocity across the multi-channel measurement history while remaining robust to measurement outliers. Within the particle filter, the coarse distance acts as an absolute range anchor that localizes the target to a coarse range window. The phase-derived relative motion refines this anchor to sub-bin precision. Its wavelength ambiguity is resolved over time by combining the anchor, the motion history, and the wavelength diversity of the two channels. The confidence score gates phase updates so that erroneous extractions do not propagate into the filter. A particle filter is used since it can represent nonlinear, multimodal, and non-Gaussian uncertainty, allowing it to maintain and resolve multiple ambiguous distance hypotheses (e.g., phase wrapping), whereas  Kalman filter variants (e.g., KF, EKF, UKF) assume a single Gaussian distribution and cannot handle such ambiguity robustly.

\subsection{Amplitude-Based Radar Pipeline}
\label{section:amplitude}
 
\begin{figure}[t]
\centering
\begin{minipage}[c]{0.5\textwidth}
\centering \includegraphics[width=\textwidth]{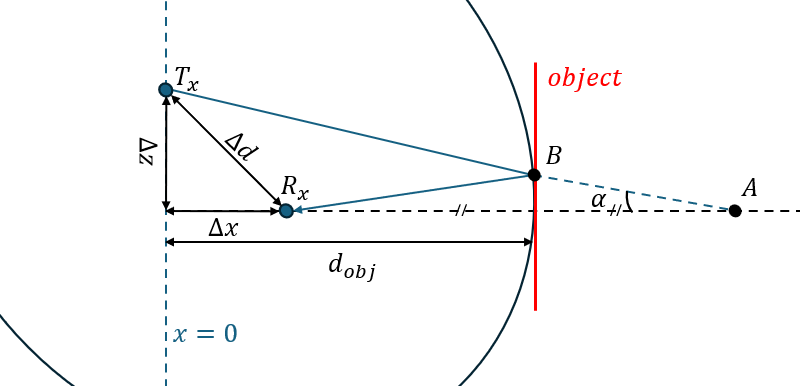}
\end{minipage}%
\caption{Geometry used for 1D distance calculation based on ToF.}
\label{fig:geometry}
\end{figure}

\noindent The radar is installed on a mobile robot in a pseudo-monostatic configuration: the transmitter $T_x$ and receiver $R_x$ are technically bistatic, but mounted close together on the same platform with a small fixed baseline. The geometry is shown in Fig.~\ref{fig:geometry}, where $\mathbf{p}_{\text{tx}}$ and $\mathbf{p}_{\text{rx}}$ denote the transmitter and receiver positions. Each \ac{ToF} measurement defines an ellipsoidal surface with these points as foci. Rather than estimating the full target position, we estimate a single scalar distance corresponding to the separation between the radar and the closest reflection point on the target. We assume that this dominant reflection originates from a known forward direction relative to the radar, and that radar or target motion occurs along this axis. To calculate this distance, the \ac{ToF} must first be extracted using signal processing.
 
Each raw CIR measurement $h_i^{(c)}[x]$ is first upsampled using spline interpolation to increase effective temporal resolution, allowing finer localization of peaks within the CIR. Candidate peaks are then identified using a \ac{CA-CFAR} thresholding approach \cite{UWB_overview}, which compares the magnitude of each sample to a locally estimated noise floor:
\begin{equation}
    |h_i^{(c)}[x]| > \beta \cdot \sigma_{\text{noise}}(x),
    \label{eq:CFAR}
\end{equation}
\noindent where $\sigma_{\text{noise}}(x)$ is computed from neighboring range bins and $\beta$ is a detection threshold. For each detected peak, a local signal-to-noise ratio $\text{SNR}_i^{(c)}$ is estimated, and the peak with the highest SNR is selected as the dominant target reflection. This selection strategy improves robustness against multipath interference and low-amplitude peaks. If no peaks satisfy Eq~\ref{eq:CFAR}, the peak of the previous timestamp $i-1$ is chosen.
 
Finally, the \ac{ToF} $\tau_i^{(c)}$ associated with the selected peak is converted into a one-dimensional distance estimate $d_i^{(c)}$ using the known bistatic radar geometry:
\begin{equation}
d_i^{(c)} = f(\tau_i^{(c)}, \mathbf{p}_{\text{tx}}, \mathbf{p}_{\text{rx}}),
\end{equation}
where the function $f(\cdot)$ converts the measured time-of-flight into a distance along the radar forward axis using the bistatic geometry shown in Fig.~\ref{fig:geometry}. In addition to the target reflection peak, the CIR consistently contains a strong early peak corresponding to the direct \ac{LOS} signal between transmitter and receiver. This peak provides a timing reference that allows absolute bistatic path lengths to be recovered.
 
Let $\tau_{\text{LOS}}$ denote the time-of-flight of the \ac{LOS} peak and $\tau_{\text{peak}}$ the time-of-flight of the selected target reflection. The total propagation length of the reflected ray is then given by
\begin{equation}
L_{\text{tot}} = \|\mathbf{p}_{\text{tx}} - \mathbf{p}_{\text{rx}}\| + c \, (\tau_{\text{peak}} - \tau_{\text{LOS}}),
\end{equation}
where $c$ is the speed of light in air. This expression accounts for the known transmitter--receiver baseline and the excess path length introduced by the reflection.
 
As illustrated in Fig.~\ref{fig:geometry}, the reflected ray can be conceptually extended to a point $A$ at the same vertical coordinate as the receiver. Because this forms a right-angled triangle, the angle $\alpha$ is computed as
\begin{equation}
\alpha = \arcsin\!\left( \frac{\Delta z}{L_{\text{tot}}} \right),
\end{equation}
where $\Delta z$ is the vertical offset of the receiver.
 
By the symmetric reflection properties of ellipsoids, the horizontal distance from the receiver to the reference point $B$ is equal to the horizontal distance from $B$ to $A$. With this knowledge, the distance of the object from $x=0$ can be expressed as:
\begin{equation}
\label{eq:distance_function}
d_\text{obj} = \frac{\Delta x + \frac{\Delta z}{\tan \alpha}}{2}.
\end{equation}
The resulting one-dimensional distance estimates form a robust initial approximation of the target position, enabling further refinement using higher-resolution phase or tracking techniques.
 
\subsection{Phase Extraction Pipeline}
\label{section:phase_pipeline}
\noindent Starting from the raw phase model of Eq~\ref{eq:raw_phase}, we cancel the offset and drift terms by subtracting the phase of the \ac{LOS} peak from the measured phase of the object reflection. The phase of the signal reflected from the object can be written as:
\begin{equation}
\begin{split}
\Phi_{\text{obj}} &= \Phi(t + \Delta t) \\
&= \Big[-2 \pi f_c (\tau_{\text{LOS}} + \Delta t)
   - 2 \pi \Delta f_c (t + \Delta t) \\
&\quad + (\Phi^{\text{tx}} - \Phi^{\text{rx}})\Big] \mod 2\pi
\end{split}
\end{equation}
 
\noindent Subtracting $\Phi_{\text{LOS}}$:
\begin{equation}
\begin{split}
\Phi_{\text{obj}} - \Phi_{\text{LOS}} &=
[-2 \pi f_c (\tau_{\text{LOS}} + \Delta t)
   - 2 \pi \Delta f_c (t + \Delta t) \\
&\quad + (\Phi^{\text{tx}} - \Phi^{\text{rx}})
   + 2 \pi f_c \tau_{\text{LOS}} \\
&\quad + 2 \pi \Delta f_c t
   - (\Phi^{\text{tx}} - \Phi^{\text{rx}})] \mod 2\pi
\end{split}
\end{equation}
which simplifies to
\begin{equation}
    \begin{split}
        \Phi_{\text{obj}} - \Phi_{\text{LOS}}  &=  [-2 \pi f_c \Delta t - 2 \pi \Delta f_c \Delta t ] \quad mod  \quad 2 \pi
        \\ &\cong [-2 \pi f_c \Delta t] \quad mod  \quad 2 \pi
        \\ &= \phi_{\text{obj}} \qquad
    \end{split}
\end{equation}
\noindent The residual \ac{CFO} term ($\Delta f_c$) can be ignored as it is typically around 50~kHz (from a $\pm$ 20 ppm oscillator \cite{CFO}) and the \ac{LOS} peak and the object peak are only separated by a few nanoseconds. For peaks 20\,ns apart, corresponding to an object distance of roughly 3\,m, the residual phase offset is
\begin{equation}
\begin{aligned}
\Delta \phi &= 2\pi \, \Delta f_c \, t \\
            &= 2\pi \times 50{,}000 \times 20 \times 10^{-9} \\
            &= 2\pi \times 10^{-3} \approx 0.0063\,\text{rad} \approx 0.36^{\circ}
\end{aligned}
\end{equation}
which can be considered negligible for high accuracy ranging.
 
To recover the fine relative motion of the target, we extract this drift-corrected phase at the target bin obtained through \ac{CFAR} in Section~\ref{section:amplitude}, and denote it $\phi_i^{(c)}$ for channel $c$ at slow time $i$. Subsequently, the target phase values of different \acp{CIR} from the same channel can be differenced to calculate a meaningful relative phase change:
\begin{equation}
    \Delta \phi_i^{(c)}=\phi_i^{(c)}-\phi_{i-1}^{(c)}
    \label{eq:phase_diff}
\end{equation}
\noindent This phase change is directly used by the particle filter, as differencing Eq~\ref{eq:phase_distance} relates it to an ambiguous relative motion through:
\begin{equation}
    \Delta d_i^{(c)}= \left(\Delta\phi_i^{(c)} / 2\pi +N\right)\lambda^{(c)},
\end{equation}
\noindent where $\Delta d_i^{(c)}$ is an ambiguous distance change and $\lambda^{(c)}$ is the carrier wavelength of channel $c$.
 
\subsection{Channel Reliability Estimation} \label{section:channel_weight}
\noindent As modeled in Section~\ref{section:channel_failure}, UWB channels experience interference and Fresnel fading at different times, which can be exploited to improve distance estimation. This is particularly critical for phase-based measurements, where extracting phase from an incorrect peak can produce entirely erroneous values. To address this, we define a confidence metric for each channel $c$ at time $i$ that quantifies the trustworthiness of its measurement. This metric allows high-confidence channels to dominate the fusion process, reducing the propagation of corrupted measurements into the refined distance estimates.
 
The SNR is obtained directly from the CFAR peak detection step in Section~\ref{section:amplitude}, which yields a local signal-to-noise ratio $\text{SNR}_i^{(c)}$ per measurement reflecting the prominence of the selected reflection relative to the noise floor. This SNR is mapped to a normalized confidence score using a sigmoid function:
\begin{equation}
w_i^{(c)} = \frac{1}{1 + \exp\!\left(-k_s^{(c)}\left(\text{SNR}_i^{(c)} - \text{SNR}_0^{(c)}\right)\right)}, \end{equation}
where $k_s^{(c)}$ and $\text{SNR}_0^{(c)}$ control the slope and midpoint per channel, and are determined empirically as shown in Section~\ref{sec:params}. This suppresses low-quality channels while retaining high-SNR measurements.
 
Channels are assigned high confidence when the detected peak is salient and originates from a strong, reliable reflection. Additional confidence metrics, such as peak characteristics (width, prominence, ...), motion consistency compared to prior measurements or inter-channel agreement, can be incorporated multiplicatively, allowing the framework to be extended without altering the fusion stage.
 
\subsection{Particle Filter Fusion}
 
To robustly combine multi-channel measurements, we employ a particle filter (PF) that approximates the posterior distribution of the target state. The PF fuses the three upstream streams: the coarse amplitude-based distance from the radar pipeline (A), the relative phase change from the phase extraction pipeline (B), and the per-channel confidence score from the reliability estimation pipeline (C).
 
\subsubsection{State Representation}
 
Each particle \(n\) at time step \(i\) represents a possible target state
\begin{equation}
\mathbf{x}_i^{(n)} =
\begin{bmatrix}
d_i^{(n)} \\
v_i^{(n)} \\
\{ d_{i-1}^{(c),(n)} \}_{c \in \mathcal{C}}
\end{bmatrix},
\end{equation}
where \(d_i^{(n)}\) and \(v_i^{(n)}\) are the particle's distance and velocity. Additionally, the state stores its previous distance per channel to enable channel-specific phase likelihood evaluation. Each particle is assigned a weight \(w_i^{(n)}\), representing the posterior probability of that state given all measurements up to time \(i\) (i.e. the measurement history):
\begin{equation}
w_i^{(n)} \propto p(\mathbf{x}_i^{(n)} \mid \mathbf{z}_{1:i}), \quad \sum_{n=1}^N w_i^{(n)} = 1,
\end{equation}
where \(\mathbf{z}_{1:i}\) denotes all measurements collected up to time \(i\).
At the first time step, all $N$ particles are randomly distributed around an initial guess of the target's distance and velocity. Each particle represents a potential state of the target, and all particles are assigned equal weight, reflecting that we have no prior reason to favor one over another.
 
\subsubsection{Prediction Step}
 
Particles are propagated according to a constant-velocity motion model with additive Gaussian noise:
\begin{align}
d_i^{(n)} &= d_{i-1}^{(n)} + v_{i-1}^{(n)} \Delta t + \epsilon_d, \quad \epsilon_d \sim \mathcal{N}(0, \sigma_d^2), \\
v_i^{(n)} &= v_{i-1}^{(n)} + \epsilon_v, \quad \epsilon_v \sim \mathcal{N}(0, \sigma_v^2),
\end{align}
where \(\Delta t\) is the time increment between steps. The constant-velocity model is commonly used in robotics and target tracking applications because it provides a simple yet effective approximation of short-term motion while requiring minimal computational resources.
 
\subsubsection{Measurement Update}
 
At each time step, the particle filter processes the measurement from a single channel. For channel $c$, the measurement vector is defined as
\begin{equation}
\mathbf{z}_i^{(c)} =
\begin{bmatrix}
d_i^{(c)} \\[1mm]
\Delta \phi_i^{(c)}
\end{bmatrix},
\end{equation}
where $d_i^{(c)}$ is the coarse distance estimate from the amplitude pipeline, and $\Delta \phi_i^{(c)}$ is the relative phase change obtained from the phase extraction pipeline.
 
The likelihood of the measurement $\mathbf{z}_i^{(c)}$ given particle $\mathbf{x}_i^{(n)}$ is factorized as
\begin{equation}
\label{eq:meas_likelihood}
p(\mathbf{z}_i^{(c)} \mid \mathbf{x}_i^{(n)}) =
p(d_i^{(c)} \mid d_i^{(n)}) \,
p(\Delta \phi_i^{(c)} \mid d_i^{(n)}, d_{i-1}^{(c),(n)})
\end{equation}
We model the coarse distance measurement likelihood as a Gaussian distribution centered on the particle's predicted distance. In practice, CFAR occasionally selects an incorrect peak, producing large non-Gaussian outliers. The particle filter handles this naturally, as its non-parametric representation makes no Gaussian assumption on the overall posterior:
\begin{equation}
\label{eq:distance_likelihood}
\begin{split}
p(d_i^{(c)} \mid d_i^{(n)})
= \qquad \qquad \\
\frac{1}{\sqrt{2 \pi} \, \sigma_\text{range}}
\exp \Bigg[
- \frac{(d_i^{(n)} - d_i^{(c)})^2}{2 \, \sigma_\text{range}^2}
\Bigg]
\end{split}
\end{equation}
To model the relative phase likelihood, we compare the measured phase change with the phase change implied by the distance increment proposed by each particle for a given channel. By operating on distance changes rather than absolute distance, this formulation does not restrict the particle velocity to sub--half-wavelength motion and therefore allows phase-ambiguous (higher) speeds to retain high probability. In combination with the coarse distance likelihood, the true distance and velocity are resolved over time. The phase error is modeled using a wrapped Gaussian distribution:
\begin{equation}
\begin{split}
p(\Delta \phi_i^{(c)} \mid d_i^{(n)}, d_{i-1}^{(c),(n)})
= \qquad \qquad \qquad \qquad  \\
\frac{1}{\sqrt{2 \pi} \, \sigma_\text{phase}} \,
\exp \Bigg[
  - \frac{\Big(
\text{wrap} \Big[
\Delta \phi_i^{(c)}
- \frac{2 \pi}{\lambda^{(c)}} \big( d_i^{(n)}-d_{i-1}^{(c),(n)} \big)
\Big] \Big)^2}{2 \, \sigma_\text{phase}^2}
\Bigg]
\end{split}
\end{equation}
 
\noindent To maximally exploit phase information, $\sigma_\text{phase}$ must be set small, making the phase likelihood highly peaked. This however introduces a failure mode: an erroneous phase measurement causes the filter to assign low weight to all physically plausible particles, collapsing the posterior onto an incorrect state. Unlike a Kalman filter, where an outlier merely biases the mean, such collapse is difficult to recover from within a short trajectory. To prevent this, we gate phase updates using the per-channel confidence score $w_i^{(c)}$ derived in Section~\ref{section:channel_weight}. A phase update is only applied when:
\begin{equation}
    w_i^{(c)} \geq w_\text{thres}
\end{equation}
\noindent discarding unreliable measurements and falling back to amplitude-only updates for that timestep.
The particle weights are updated multiplicatively for the current channel:
\begin{equation}
w_i^{(n)} \propto w_{i-1}^{(n)} \,
p(\mathbf{z}_i^{(c)} \mid \mathbf{x}_i^{(n)}),
\end{equation}
and normalized so that
\begin{equation}
\sum_{n=1}^N w_i^{(n)} = 1.
\end{equation}
 
\subsubsection{Resampling}
 
Resampling is triggered when the effective sample size
\begin{equation}
\text{ESS} = \frac{1}{\sum_{n=1}^N (w_i^{(n)})^2}
\end{equation}
falls below a predefined threshold $N_\text{thresh}$. Systematic resampling is used to redraw particles according to their weights.
 
\subsubsection{State Estimation}
 
The posterior mean of the particles provides the current estimates of distance and velocity:
\begin{equation}
\hat{d}_i = \sum_{n=1}^N w_i^{(n)} d_i^{(n)}, \quad
\hat{v}_i = \sum_{n=1}^N w_i^{(n)} v_i^{(n)}.
\end{equation}
 
We can use $\hat{d}_i$ as our final refined distance estimate for evaluation.
 
\subsubsection{Sequential Multi-Channel Updates}
Because channels are sampled in an interleaved manner, their measurements are incorporated one at a time as they arrive. Each channel's update multiplies into the running particle weights via the likelihood of Eq.~\eqref{eq:meas_likelihood}, so a confident channel can correct the hypotheses favored by a degraded one within the same trajectory. This sequential fusion is what makes the filter robust to per-channel failures.

\section{System Implementation}
\label{sec:implementation}

This section describes the practical implementation of the proposed passive, phase-based UWB ranging system, including hardware setup, data collection, parameter choices, and measurement procedures.
\begin{figure}[t] 

\centering 
\begin{minipage}[c]{0.25\textwidth}
    \centering
    \includegraphics[width=\textwidth]{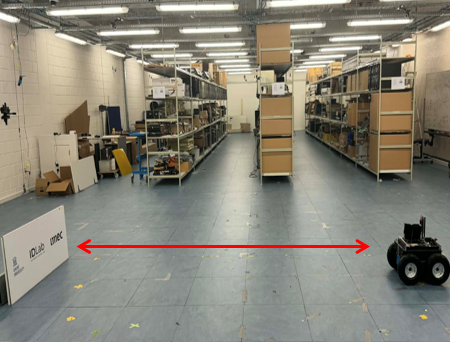}
    \end{minipage}%
\begin{minipage}[c]{0.25\textwidth} 
\centering \includegraphics[width=\textwidth]{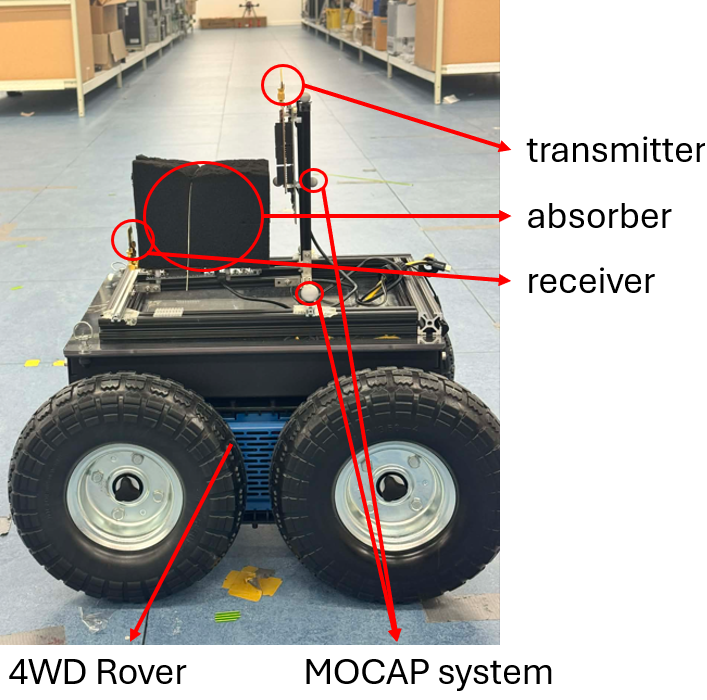} 
\end{minipage}%
\caption{Experimental setup at IIOT lab Ghent. Left: industrial indoor environment with the rover and metalic plate target. Right: bistatic radar module showing the transmitter, foam absorber, and receiver.} 
\label{fig:setup} 
\end{figure}

\subsection{Hardware Setup }

Fig.~\ref{fig:setup} illustrates the experimental setup used for evaluation. A pseudo-monostatic UWB radar system using two DW3000 transceivers \cite{qorvo_dw3000_datasheet}
was mounted on a 4WD mobile rover platform \cite{rover2026}, with the transmitter and receiver antennas separated by a baseline of 18.6 cm horizontal and 18.1 cm vertical. The DW3000 supports
UWB channels 5 and 9, with center frequencies of 6489.6
MHz and 7987.2 MHz and a bandwidth of approximately 499.2 MHz per channel. The rover was 
chosen for its ability to achieve the high forward and 
backward speeds required to evaluate system performance 
across a higher velocity range, while also reflecting a 
realistic deployment scenario for AGV navigation and SLAM 
applications. Experiments were conducted in the industrial IoT lab at IDLab Ghent University~\cite{idlab2026}, a warehouse- and factory-like environment with metal shelving, a highly reflective floor, and scattered obstacles, resulting in significant multipath effects.The ground truth of the obstacles and the robot were recorded 
at 100\,Hz using a millimeter-accurate MOCAP system 
present in the IIoT lab.

\subsection{Data collection}

All experiments were conducted in the industrial IoT lab~\cite{idlab2026} 
under controlled \ac{LOS} conditions. A flat steel plate was chosen as the target 
to isolate algorithmic performance from target-dependent variability: its 
planar geometry simplifies the 1D range conversion, and its metallic surface 
produces a strong, consistent reflection that serves as a controlled reference 
for evaluation. It is also practically motivated, as flat vertical surfaces 
such as walls and panels are the primary obstacles in \ac{AGV} and SLAM environments.

CIRs were captured sequentially on UWB channels 5 and 9 at a combined rate of 125\,Hz (62.5\,Hz per channel), with each CIR spanning 50 range bins. The dataset consists of 8 trajectories 
covering distances from 1.29 to 5.65\,m and speeds up to 1.7\,m/s, 
representative of typical AGV operations. One dedicated trajectory was designed to capture 
fine-grained displacements of 5--20\,cm to specifically evaluate phase 
sensitivity to sub-bin motion. An additional static trial was conducted 
separately to evaluate phase drift cancellation (Section~\ref{section:eval_phase}) 
and is excluded from the ranging evaluation, as is the first moving second 
of each trajectory to allow the particle filter to converge to a stable estimate.

\subsection{Parameter Choices}
\label{sec:params}

All system parameters are listed in Table~\ref{tab:params}. CFAR and CIR preprocessing 
parameters were selected based on preliminary experiments and were not further tuned 
on the evaluation dataset. Spline interpolation was chosen over linear and FFT-based alternatives based on lowest coarse ranging error in those trials.
The \ac{PF}'s particles were initialised around the first coarse distance estimate with zero mean velocity, with spread controlled by $\sigma_{p,0}$ and $\sigma_{v,0}$ respectively. The ESS resampling threshold was set to $0.5 \times N$, following standard practice in particle filtering~\cite{doucet_tutorial_nodate}. The remaining \ac{PF}-parameters were obtained through random search on a single trajectory and kept fixed across all evaluation scenarios.

\begin{figure}[t]
    \centering
    \includegraphics[width=\columnwidth]{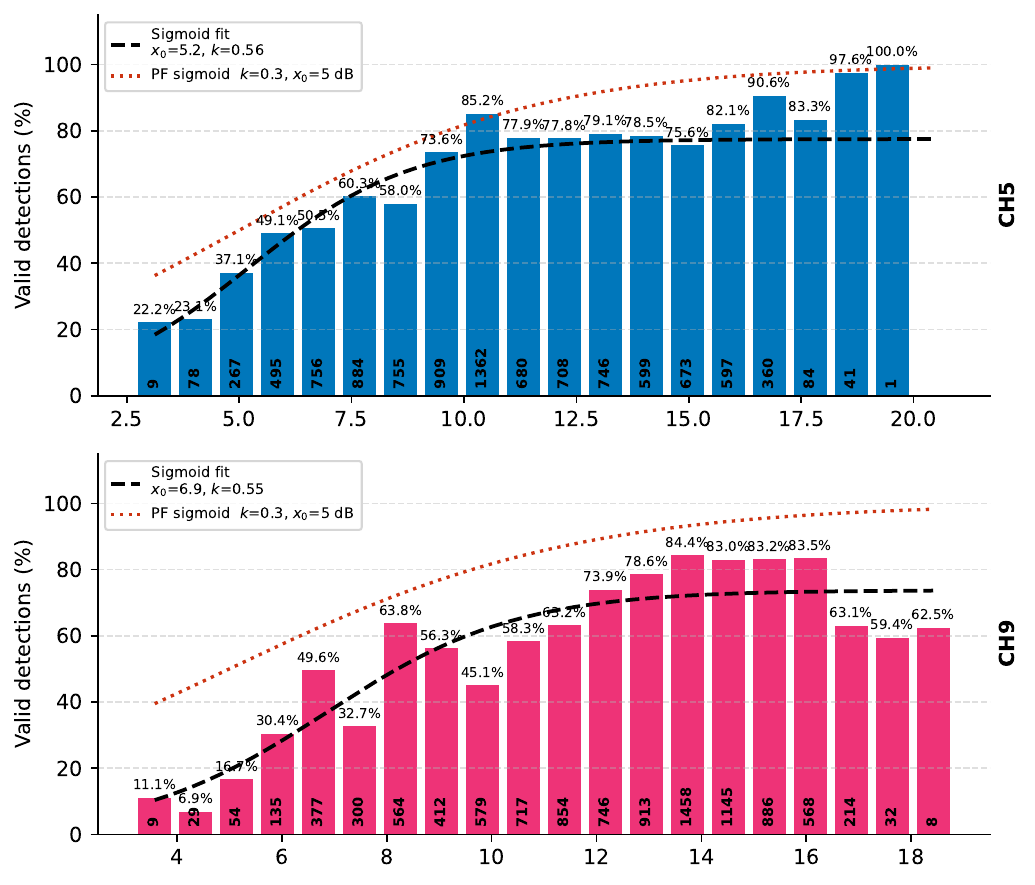}
    \caption{Fraction of valid CFAR detections (peaks within 10~cm of ground 
truth) as a function of SNR for channels~5 and~9. The dashed black curve 
shows the per-channel fitted sigmoid and the dotted red curve the 
adopted PF confidence assignment. Sample counts per bin are shown inside 
each bar.}
    \label{fig:snr_valid}
\end{figure}

The SNR sigmoid parameters and phase confidence threshold $w_{\text{thres}}$ 
were set based on the observed relationship between SNR and CFAR detection 
reliability shown in Fig.~\ref{fig:snr_valid}. The fitted sigmoid (black) 
provides a data-driven reference per channel, but using it directly would 
couple the confidence model to the specific noise conditions of the 
evaluation dataset , limiting generalization. We therefore adopt the conservative fixed curve (red), shared across both channels to avoid channel-specific overfitting. The 
threshold $w_{\text{thres}} = 0.7$ corresponds to approximately 8~dB, 
where valid detection rates remain below 60\% for both channels, meaning 
a substantial fraction of peaks are still incorrectly selected. Below this 
point, phase measurements are suppressed entirely, since they will often incorrectly and disproportionately distort particle weights due to the small $\sigma_{\text{phase}}$  .

\begin{table}[h]
\centering
\caption{System parameter values}
\label{tab:params}
\setlength{\tabcolsep}{4pt}
\renewcommand{\arraystretch}{1.1}
\begin{tabular}{|l|l|c|l|}
\hline
\textbf{Group} & \textbf{Parameter} & \textbf{Symbol} & \textbf{Value} \\
\noalign{\hrule height 1.5pt}
\multirow{2}{*}{CIR preprocessing}
    & Interpolation method      & ---                       & spline \\
    & Upsampling factor         & ---                       & $4\times$ \\
\hline
\multirow{4}{*}{CFAR}
    & Guard cells               & ---                       & 1 \\
    & Reference cells           & ---                       & 6 \\
    & Detection threshold       & $\alpha$                  & 1.7 \\
    & Smoothing                 & ---                       & 1 \\
\hline
\multirow{8}{*}{Particle filter}
    & Number of particles       & $N$                       & 1000 \\
    & Range meas. noise         & $\sigma_{\text{range}}$   & 0.07\,m \\
    & Phase meas. noise         & $\sigma_{\text{phase}}$   & 0.04\,rad \\
    & Process noise (position)  & $\sigma_{p}$              & 0.0012\,m \\
    & Process noise (velocity)  & $\sigma_{v}$              & 0.06\,m/s \\
    & Initial position std.     & $\sigma_{p,0}$            & 0.2\,m \\
    & Initial velocity std.     & $\sigma_{v,0}$            & 0.05\,m/s \\
    & ESS resampling threshold  & $N_{\text{thresh}}$       & $0.5N$ \\
\hline
\multirow{3}{*}{Channel reliability}
    & Phase conf. threshold     & $w_{\text{thres}}$        & 0.7 \\
    & Slope conf. function      & $k_s^{5,9}$               & 0.3 \\
    & Midpoint conf. function   & $\mathrm{SNR}_0^{5,9}$    & 5 \\
\hline
\end{tabular}
\end{table}

\section{Evaluation}
\label{sec:evaluation}

This section evaluates the performance of the proposed passive, phase-based UWB 
ranging system. We assess phase drift cancellation (Section~\ref{section:eval_phase}), incremental pipeline contribution (Section~\ref{section:baseline}), fine motion sensitivity (Section~\ref{section:fine_motion}), the impact of multi-channel fusion Section~\ref{section:impact_channel_fusion}), and speed robustness beyond the phase ambiguity limit (Section~\ref{section:speed}). All particle filter configurations are run 20 times per trajectory to 
average out stochastic variation inherent to the algorithm, with error 
statistics computed over all runs pooled. The first moving second of 
each trajectory is excluded to allow filter convergence.

\subsection{Phase Drift Cancellation Performance}
\label{section:eval_phase}

To evaluate phase drift cancellation, we conducted a static experiment over 55 seconds with a stationary target at CIR bin 28. This duration is sufficient to observe oscillator-induced drift, which accumulates continuously over time. Since the target does not move, the true inter-sample phase difference should be zero, and any observed deviation reflects residual oscillator-induced drift or offsets .

\begin{figure*}[t]
    \centering
    \begin{minipage}[c]{0.6\textwidth}
        \centering
        \includegraphics[width=\textwidth]{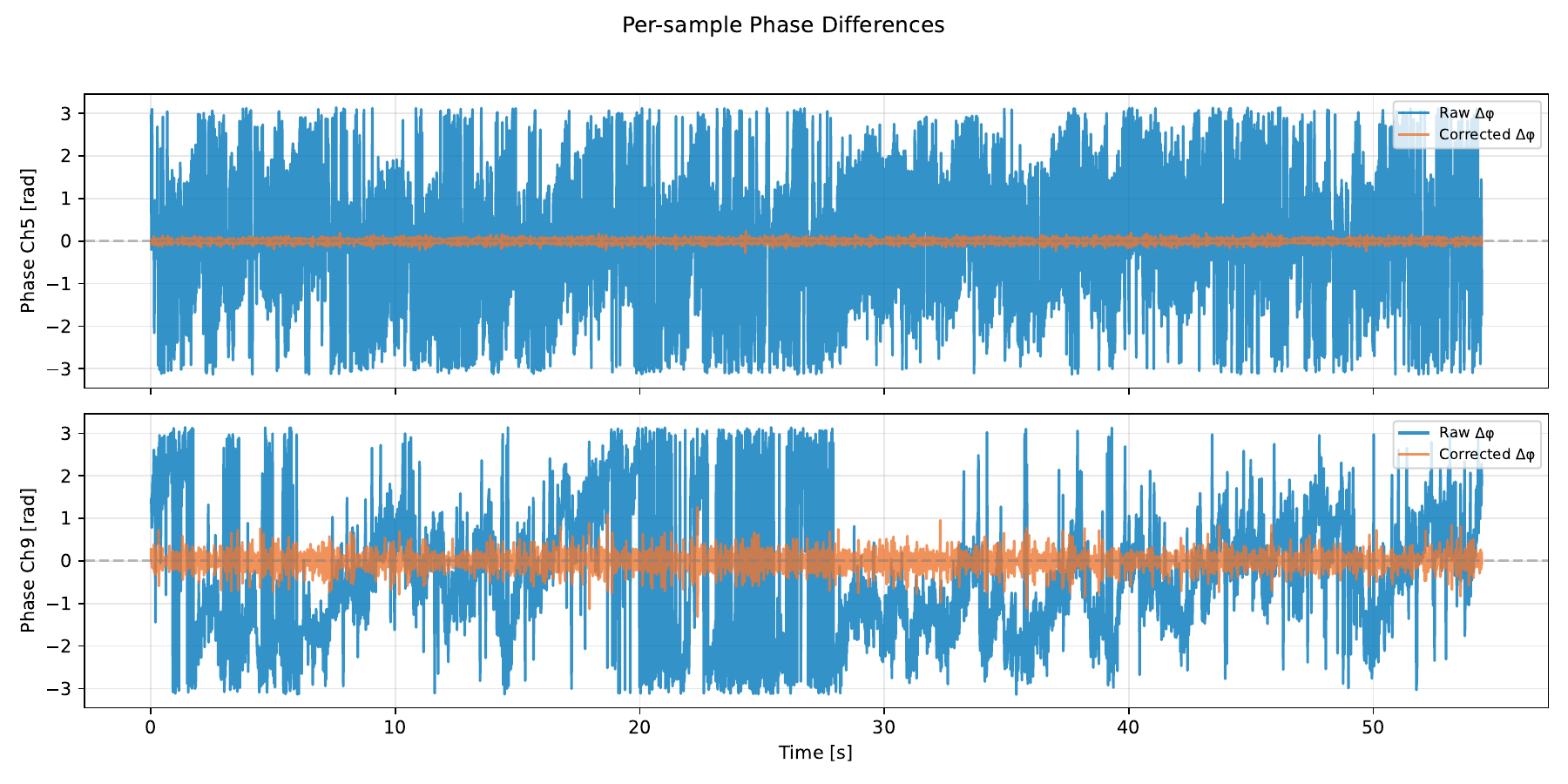}
    \end{minipage}%
    \hspace{0.02\textwidth}%
    \begin{minipage}[c]{0.35\textwidth}
        \centering
        \includegraphics[width=\textwidth]{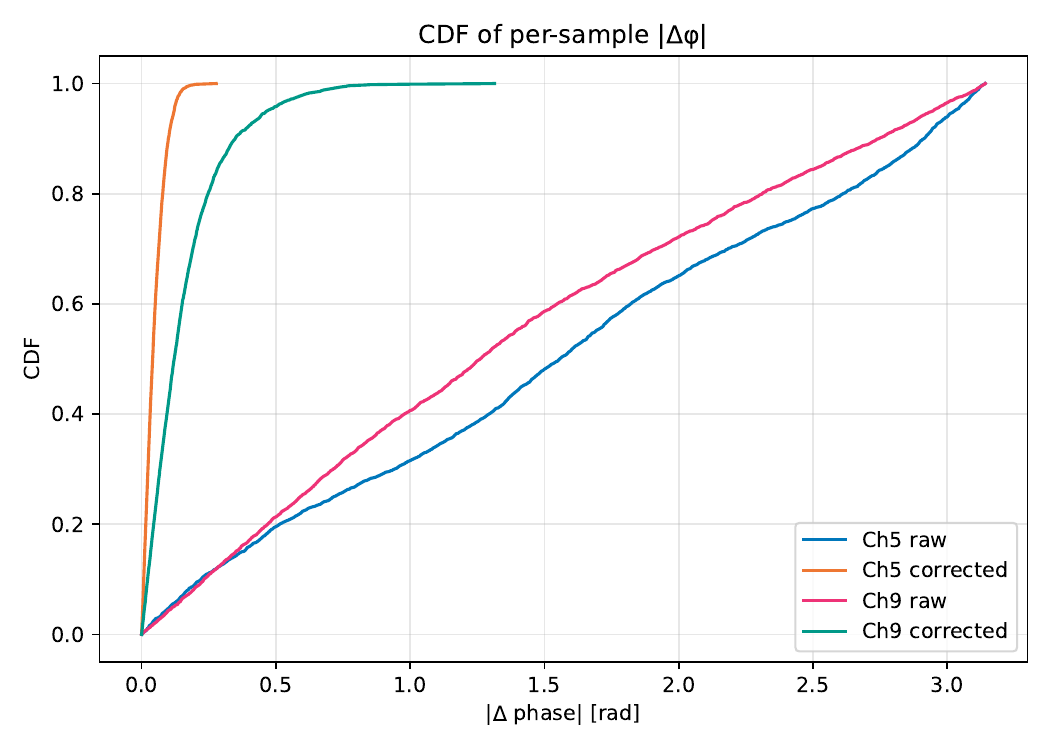}
    \end{minipage}
    \caption{{Phase drift cancellation evaluation. Left: inter-sample phase difference over time for both channels, before and after correction. Right: CDF of $|\Delta \phi|$, showing the corrected phase is sharply concentrated near zero compared to the uncorrected phase.}}
    \label{fig:phase_cdf}
\end{figure*}

Fig.~\ref{fig:phase_cdf} shows the inter-sample phase difference over time and the CDF of $|\Delta \phi|$ for both channels, before and after correction. The raw phase is effectively uniformly distributed across $[-\pi, \pi]$, confirming that oscillator-induced drift and offsets completely dominate the uncorrected measurements and renders them unusable. After correction, the phase is sharply concentrated near zero for both channels. For Channel~5, the corrected median $|\Delta \phi|$ is 0.041\,rad with a 90th percentile of 0.100\,rad. For Channel~9, the corrected median is 0.121\,rad with a 90th percentile of 0.349\,rad. These results confirm that the proposed drift cancellation successfully stabilizes the phase, making it suitable for passive UWB radar.

\subsection{Baseline Pipeline Comparison}
\label{section:baseline}
To isolate the contribution of each component, four configurations are evaluated incrementally, corresponding to pipeline stages A to D shown in Fig.~\ref{fig:system overview}: the CFAR amplitude baseline (A), particle filter fusion without phase information (A+D), phase-based fusion (A+B+D), and the full system including the gating based on channel reliability (A+B+C+D). The CFAR pipeline serves as the baseline, as CFAR-based peak detection is the conventional approach for passive UWB ranging and forms the foundation of our system. To ensure a fair comparison, the same interpolation and CFAR parameters are used across all configurations.

Table~\ref{tab:baseline} shows the ranging error statistics aggregated across all trajectories. The CFAR baseline achieves a MedAE of 6.09~cm but suffers from large outliers, reflected in the 40.01~cm standard deviation. Adding postprocessing of the CFAR baseline through the \ac{PF}, which fuses both channels' coarse estimates over time, reduces the MedAE to 2.34\,cm and standard deviation to 3.89\,cm. The PF with multi-channel phase fusion further reduces the MedAE to 1.81~cm, confirming the contribution of fine-grained phase information. Including reliability scores as a gating mechanism further reduces the error down to 1.69~cm MedAE. More importantly, the primary benefit of channel reliability weighting is robustness to outliers caused by low-SNR conditions, significantly reducing the standard deviation from 6.52~cm to 3.39~cm. The marginal $AE_{90}$ increase (4.45~cm to 4.76~cm) is the cost of this protection: gating discards some useful phase updates along with the rare but extreme outliers.

To compare these results with prior work, Van Herbruggen et al.~\cite{CoarseDistance} report 9~cm MAE on metal targets, but as discussed in Section~\ref{sec:related}, their ACIR method requires a stationary target, making a direct comparison on our moving trajectories unfair. For Giurea et al.~\cite{Adelina}, we recreated their peak detection step and evaluate it on our dataset, omitting their AoA-based filtering stages which are unavailable in our single-antenna receiver. Parameters were tuned to our configuration. Under these conditions, their pipeline achieves a MedAE of 8.44~cm and MAE of 16.97~cm, compared to 1.69~cm and 2.48~cm for our full system.

\begin{table}[h]
\centering
\caption{Ranging error statistics (cm) per pipeline configuration}
\label{tab:baseline}
\begin{tabular}{|l|c|c|c|c|}
\hline
\textbf{Configuration} & \textbf{MAE} & \textbf{MedAE} &  \textbf{AE\textsubscript{90}} &  \textbf{Std} \\
\noalign{\hrule height 1.5pt}

Giurea et al. \cite{Adelina}   & 16.97 & 8.44  & 29.11   & 36.93 \\
\hline
CFAR Baseline (A)                  & 16.11 & 6.09  & 27.21  & 40.01 \\
\hline
PF (A+D)                        & 3.27  & 2.34  & 6.60    & 3.89  \\
\hline
PF + Phase (A+B+D)                     & 2.70  & 1.81  & \textbf{4.45}   & 6.54  \\
\hline
PF + Phase + Gating (A+B+C+D)        & \textbf{2.48}  & \textbf{1.69}    & 4.76  & \textbf{3.39}  \\
\hline
\end{tabular}
\end{table}

\subsection{Phase Sensitivity Under Fine-Grained Motion}
\label{section:fine_motion}

To further highlight the contribution of phase, we isolate a dedicated fine-grained motion
trajectory with small irregular back-and-forth displacements($1$--$20$~cm) , a harder
scenario where the particle filter's constant-velocity motion model provides less
predictive power and phase differences are the primary source of sub-bin accuracy.

\begin{figure}[t]
    \centering
    \begin{minipage}[c]{0.5\textwidth}
        \centering
        \includegraphics[width=\textwidth]{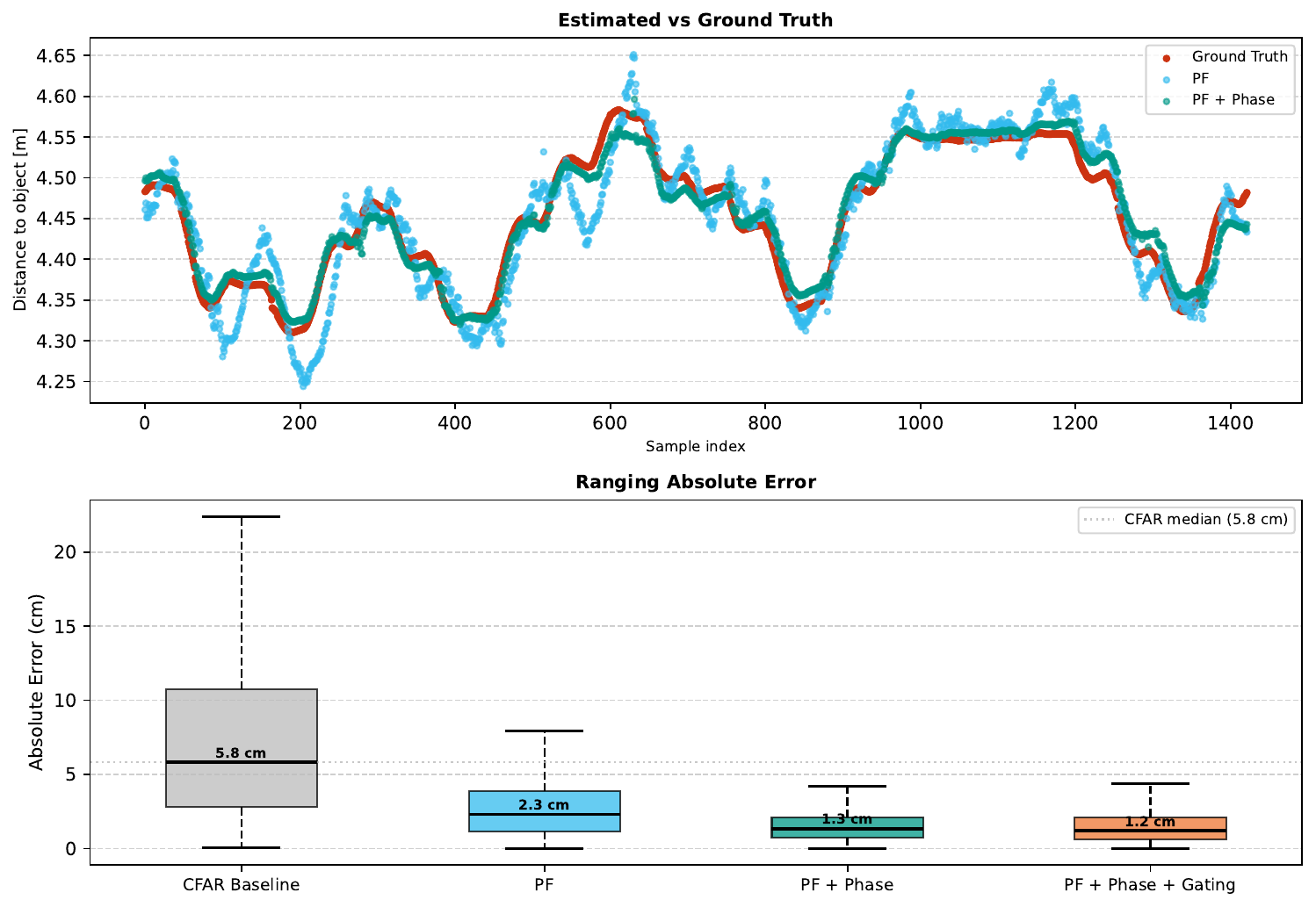}
    \end{minipage}%
    \caption{Fine-grained motion trajectory evaluation with irregular sub-wavelength movements. Top: estimated versus ground truth 
    distance for the amplitude-only pipeline (PF) and the phase-augmented pipeline 
    (PF + Phase), showing that adding phase information yields closer tracking of 
    irregular motion. Bottom: ranging absolute error comparison across all pipeline 
    configurations on the same trajectory. Median values are annotated per box.}
    \label{fig:results_boxplots}
\end{figure}

Fig.~\ref{fig:results_boxplots} shows that adding phase information nearly halves the
PF MedAE from $2.3$~cm to $1.3$~cm ($43\%$) on this trajectory, compared to a $23\%$
reduction in the general case (from $2.35$~cm to $1.81$~cm). The full system including
channel reliability weighting further reduces MedAE to $1.2$~cm, demonstrating that
both phase and confidence gating contribute meaningfully when motion is unpredictable
and the amplitude-only estimate is least reliable. This is further confirmed
qualitatively by the trajectory plots: the full pipeline closely tracks the irregular
ground truth motion, while the amplitude-only estimate is visibly noisier and less
responsive to direction changes.

\subsection{Impact of Multi-Channel Fusion}
\label{section:impact_channel_fusion}
The impact of multi-channel fusion is evaluated by comparing single-channel 
and dual-channel configurations of the full system (A+B+C+D). Since individual 
UWB channels can exhibit distinct SNR characteristics, as illustrated in 
Fig.~\ref{fig:interference}, fusing two channels is expected to improve 
robustness against channel-specific degradation. To ensure a fair comparison, 
dual-channel configurations are downsampled to 62.5\,Hz, matching the 
per-channel sampling rate of single-channel configurations.

\begin{figure}[t]
    \centering
    \begin{minipage}[c]{0.5\textwidth}
        \centering
        \includegraphics[width=\textwidth]{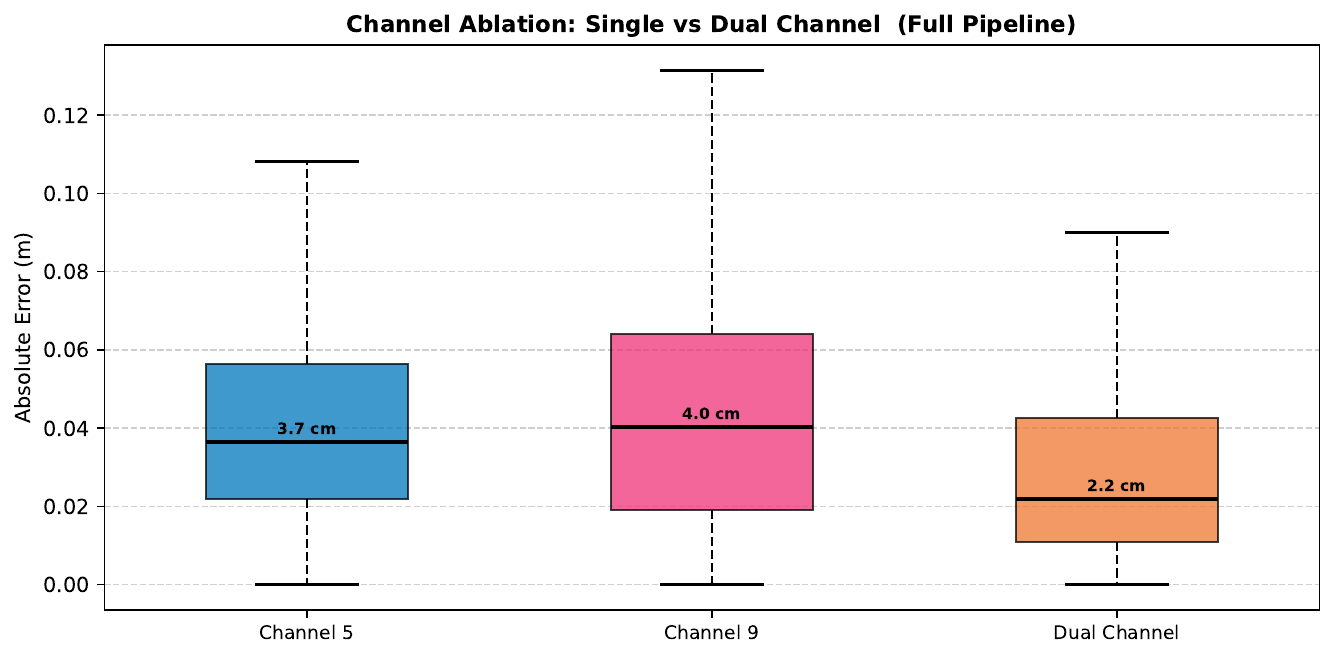}
    \end{minipage}%
    
    \caption{Ranging absolute error for single-channel and dual-channel 
configurations using the full pipeline. Median values are annotated 
per box. }
    \label{fig:channel_ablation}
\end{figure}

As shown in Fig.~\ref{fig:channel_ablation}, dual-channel fusion achieves a 
MedAE of $2.2$\,cm compared to $3.7$\,cm and $4.0$\,cm for CH5 and CH9 
respectively. Notably, CH9 performs worse than CH5, consistent with its higher 
noise floor observed in Fig.~\ref{fig:interference}, further motivating the 
channel reliability weighting component of our system.

\subsection{Speed Robustness Beyond the Phase Ambiguity Limit}
\label{section:speed}

Next, we evaluate whether the proposed system maintains ranging accuracy at speeds exceeding the per-channel phase ambiguity limits derived in Section~\ref{section:pbr}. From Eq~\ref{eq:vmax}, the theoretical limit is $0.72$~m/s for Channel~5 and $0.59$~m/s for Channel~9 at $62.5$~Hz per-channel sampling rate. Above these limits, single channel phase information cannot unambiguously track inter-sample motion. Robustness is instead expected from the \ac{PF} jointly  resolving ambiguity through the coarse amplitude anchor and multi-channel phase info. 

\begin{figure}[t]
    \centering
    \begin{minipage}[c]{0.5\textwidth}
        \centering
        \includegraphics[width=\textwidth]{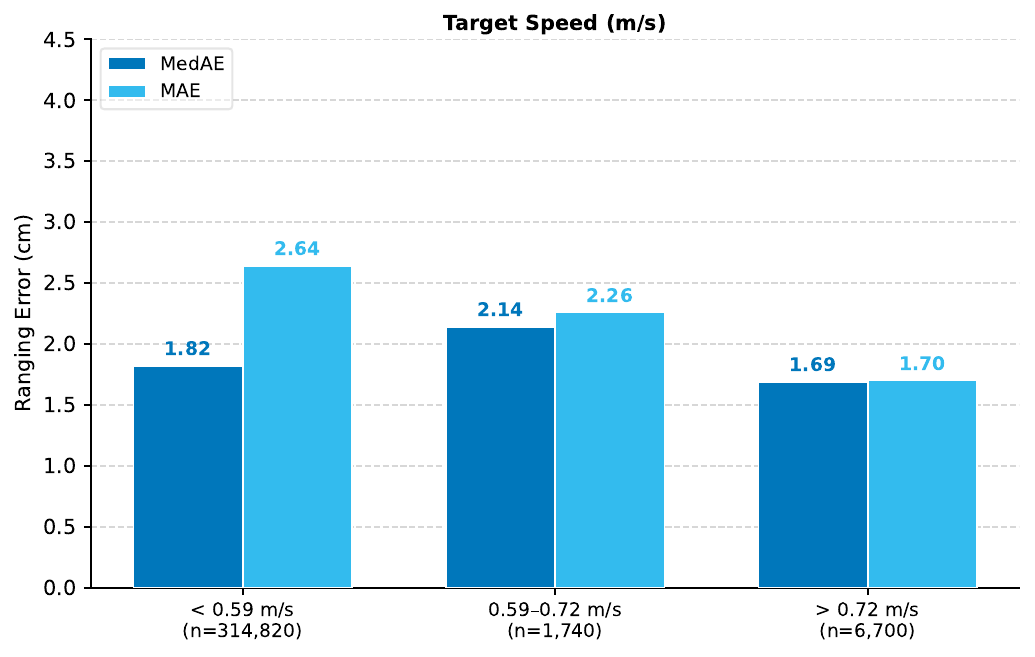}
    \end{minipage}%
    
    \caption{Distance estimation accuracy across different target speed regimes, segmented based on the phase ambiguity limits of individual channels. The proposed multi-channel fusion approach maintains consistent performance even when one or both channels become ambiguous, highlighting its robustness to phase wrapping. 
    }
    \label{fig:speed}
\end{figure}

To assess this, all trajectory samples are segmented by instantaneous speed obtained from the MOCAP ground truth into three bins: below $0.59$~m/s (unambiguous), between $0.59$ and $0.72$~m/s (ambiguous for Channel~9 only), and above $0.72$~m/s (ambiguous for both channels). As shown in Fig.~\ref{fig:speed}, MedAE remains below $2.2$~cm across all bins, with 
no degradation observed above either ambiguity limit.

\begin{figure}[t]
    \centering
    \begin{minipage}[c]{0.5\textwidth}
        \centering
        \includegraphics[width=\textwidth]{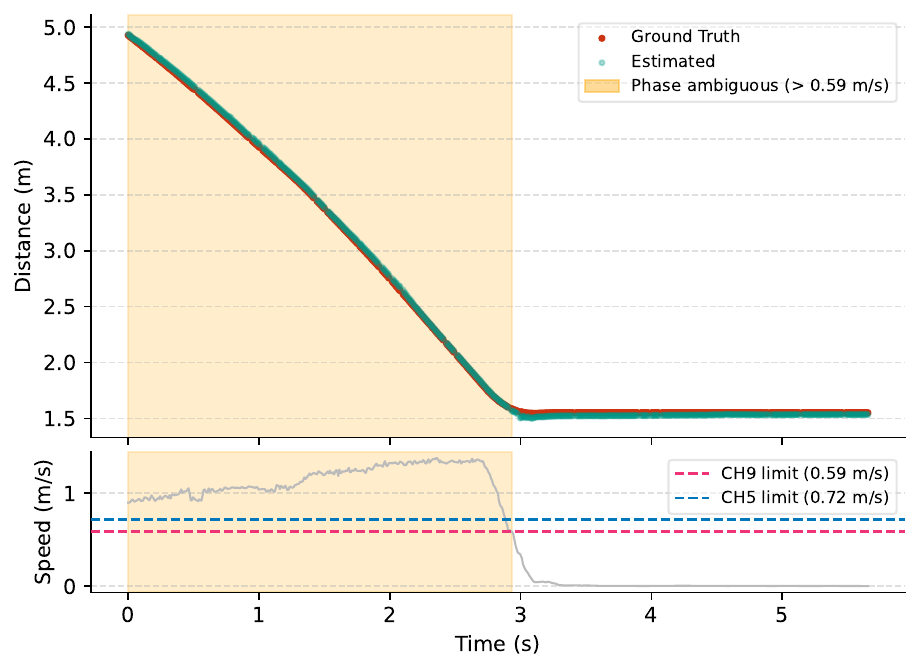}
    \end{minipage}%
    
    \caption{Estimated versus ground truth distance for the highest-speed trajectory (single particle filter run shown for clarity), with instantaneous speed profile below. The shaded region indicates segments exceeding the CH9 phase ambiguity limit ($0.59$~m/s).}
    \label{fig:speed_traj}
\end{figure}

Fig.~\ref{fig:speed_traj} further illustrates the behavior of the system on the highest-speed trajectory, reaching $1.7$~m/s. The estimated distance closely tracks the MOCAP ground truth throughout, including segments that consistently exceed both ambiguity limits. Within this phase-ambiguous region, the system achieves a MedAE of $1.68$~cm, MAE of $1.69$~cm, and AE90 of $2.97$~cm across 20 particle filter runs, demonstrating that the particle filter maintains stable and accurate ranging under phase-ambiguous motion.

\subsection{Discussion}
\label{section:discussion}
\noindent Phase information is most valuable precisely in scenarios where the motion model provides limited guidance. For small-movement trajectories or trajectories with frequent direction reversals, the constant-velocity assumption offers little predictive power, leaving phase as the primary source of sub-bin information. This explains the disproportionate accuracy improvement observed in these cases ($48\%$), compared to the more moderate gain in the general case ($23\%$). The fact that phase carries most of the signal in this regime also suggests that further accuracy gains are most likely to come from better exploiting phase information under unpredictable motion, with concrete directions discussed in Section~\ref{sec:future_work}.

At first glance, Table~\ref{tab:baseline} suggests the particle filter is doing most of the work: MedAE drops from 6.09\,cm at the CFAR baseline to 2.34\,cm once the PF is added. However, Section~\ref{section:impact_channel_fusion} shows that single-channel operation under the full pipeline still sits at 3.7--4.0\,cm, so a substantial share of the gain comes from fusing two channels' coarse estimates rather than from temporal smoothing alone. This holds even when one channel has a higher noise floor, because the channels' impairments are only partially correlated. The key design implication is that spectral diversity matters more than per-channel quality, motivating acquisition strategies that maximize frequency separation.

Beyond accuracy, the PF and reliability gating together provide the system's robustness. The CFAR baseline is occasionally catastrophically wrong, with its standard deviation roughly $7\times$ its median error. The PF absorbs these failures into the posterior, and gating extends this further by suppressing corrupted phase measurements before the sharp phase likelihood can collapse the \ac{PF} onto an incorrect state. This matters most in safety-critical AGV deployment, where outliers drive system requirements.

\section{Future research directions}
\label{sec:future_work}

\noindent While our results significantly outperform prior work, achieving millimeter-level accuracy in passive UWB radar in realistic environments requires addressing several limitations of the current system design. This section first discusses the limitations of the current system and then outlines extensions addressing these challenges.

\begin{itemize}

\item First, all experiments were conducted using a flat metallic target. Extending the approach to more complex materials and geometries requires more discriminative peak selection and channel reliability mechanisms beyond \ac{CFAR} and SNR-based gating. In particular, richer peak descriptors such as peak prominence, width, temporal stability, phase consistency, and inter-channel agreement could improve robustness under dense multipath conditions. Instead of hard phase gating, a probabilistic weighting scheme for unreliable measurements may further improve stability in cluttered environments.

\item Further improvements may be achieved through polarization-based peak filtering, which provides an additional degree of discrimination between reflection mechanisms. Specular reflections tend to preserve polarization more strongly than diffuse or multi-bounce components, enabling improved suppression of spurious multipath peaks. Reducing such contributions would also yield a cleaner phase at the target bin, thereby lowering phase noise and improving ranging accuracy. Beyond peak filtering, polarization information may also provide additional cues about surface orientation and material properties, further enhancing clutter rejection.

\item A natural extension is joint range--angle estimation, where both distance and direction are inferred within a unified fusion framework. Multi-antenna configurations would enable angle-of-arrival estimation for 2D tracking. In this setting, reflections from implausible directions can be rejected, while temporal consistency in angle estimates can serve as an additional reliability constraint for suppressing spurious paths.

\item The current framework is limited to single-target estimation. Extending it to multi-object tracking requires robust data association of reflection peaks over time, reinforcing temporally consistent peaks while suppressing transient ones. This association is particularly critical in phase-based systems, since phase differencing is only meaningful when consecutive observations correspond to the same physical target, motivating tracking strategies with explicit identity preservation.

\item Finally, channel acquisition is currently sequential since the DW3000 supports only one active channel at a time, introducing small temporal offsets between per-channel measurements. Hardware supporting simultaneous multi-channel acquisition would enable synchronous CIR observations across channels, improving fusion accuracy and reliability under higher target dynamics.

\end{itemize}

\noindent Taken together, these directions outline a path toward robust, millimeter-level, tag-free UWB sensing in complex real-world environments.

\section{Conclusion}
\label{sec:conclusion}
\noindent Passive UWB radar on commercial hardware has long been limited by amplitude-based processing, with the best reported result at 9\,cm MAE. Phase-based methods have broken this barrier in active tag-based systems, but in passive radar reflected signals arrive weaker and mixed with multipath, and commercial hardware provides high-resolution timing only for the first path. Phase extraction is further complicated by oscillator drift, inherent phase ambiguity, and mixed multipath at the target bin, leaving tagless phase-based ranging largely unaddressed.

This work closes that gap. By cancelling oscillator drift against the \ac{LOS} peak and fusing reliability-weighted relative phase with coarse amplitude estimates in a multi-channel particle filter, the system achieves a median error of 1.69\,cm on a DW3000 bistatic radar and a 3.6$\times$ improvement over the best prior result. Beyond accuracy, the framework explicitly accounts for channel unreliability and phase corruption, enabling stable operation under channel-specific degradation and high-speed motion. It improves MedAE by more than 40\% over single-channel operation and maintains accuracy at speeds beyond both channels' phase ambiguity limits.

While the current system achieves consistent centimeter-level performance, a roadmap toward sub-centimeter accuracy is provided through further innovations in peak selection, channel reliability estimation, polarization-based filtering, multi-dimensional sensing (e.g., joint range--angle estimation), and multi-target tracking.




\bibliographystyle{IEEEtran}
\bibliography{reference.bib}

\newpage

 




\vfill

\end{document}